\documentclass{PoS}

\usepackage{xspace}

\usepackage{amsmath}

\usepackage{gensymb}

\usepackage{float}

\usepackage[hypcap]{caption}

\usepackage{lineno}
\label{mathrefs}

\usepackage{subcaption}

\usepackage{graphicx}

\def \eeflux   {$e^-+e^+$\xspace}
\def \g {gamma\xspace}
\def \gh {gamma/hadron\xspace}

\title{Gamma Hadron Separation using Pairwise Compactness Method with HAWC}

\ShortTitle{HAWC Pairwise Compactness}
\author{\speaker{Zigfried Hampel-Arias}$^a$ and Stefan Westerhoff$^a$ for the HAWC Collaboration$^b$ \\
        \llap{$^a$}Department of Physics and WIPAC, University of Wisconsin-Madison, Madison, WI, USA \\
        \llap{$^b$}For a complete author list, see \href{http://www.hawc-observatory.org/collaboration/icrc2015.php}{www.hawc-observatory.org/collaboration/icrc2015.php}.

Email: \email{zhampel@icecube.wisc.edu}}

\abstract{
The High-Altitude Water Cherenkov (HAWC) Observatory is a ground based air-shower array deployed on the slopes of Volcan Sierra Negra in the state of Puebla, Mexico. While HAWC is optimized for the detection of gamma-ray induced air-showers, the background flux of hadronic cosmic-rays is four orders of magnitude greater, making background rejection paramount for gamma-ray observations. On average, gamma-ray and cosmic-ray showers are characterized by different topologies at ground level. We will present a method to identify the primary particle type in an air-shower that uses the spatial relationship of triggered PMTs (or "hits") in the detector. For a given event hit-pattern on the HAWC array, we calculate the mean separation distance of the hits for a subset of hit pairs weighted by their charges. By comparing the mean charge and mean separating distance for the selected hits, we infer the identity of the event's primary. We will report on the efficiency for identifying gamma-rays and the performance of the technique with simulation.}

\FullConference{The 34th International Cosmic Ray Conference,\\
		30 July- 6 August, 2015\\
		The Hague, The Netherlands}

\begin{document}

\section{Introduction}
Gamma-ray observations at GeV-TeV energies are obtained by ground-based experiments in two categories. Imaging air-Cherenkov telescopes (IACTs), including VERITAS, MAGIC, and H.E.S.S., have a low duty cycle (\char`\~10\%) but excellent sensitivity over a narrow field-of-view (<5\degree). Air-shower arrays, including Milagro, ARGO-YBJ, and HAWC, have comparatively lower instantaneous sensitivity but have a nearly continuous duty cycle and wide field-of-view (\char`\~$2 sr$) of the overhead sky. This advantage gives air-shower arrays the ability to make unbiased surveys of the GeV-TeV \g-ray sky, and will complement IACT observations at higher energies (up to 100 TeV), with greater sensitivity for extended sources.

The challenge for air-shower arrays to observe \g-ray sources 
is the large background of hadronic cosmic rays. Thus, the identification and rejection 
of this background is paramount. Fortunately, the nature of the primary particle determines 
the subsequent shower development, and provides air-shower arrays such as 
HAWC a manner in which to discriminate event types.

\begin{figure}[H]
\centering

\begin{subfigure}[t]{0.45\textwidth}
  \centering
  \includegraphics[width=1\linewidth]{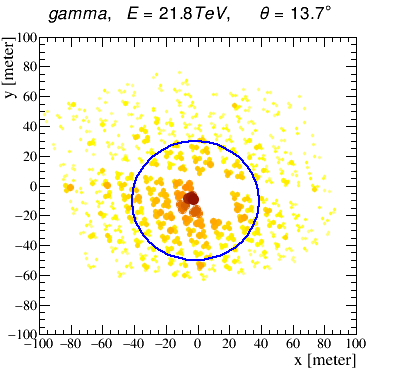}
  \caption{}
  \label{fig:gevent}
\end{subfigure}
~
\begin{subfigure}[t]{0.45\textwidth}
  \centering
  \includegraphics[width=1\linewidth]{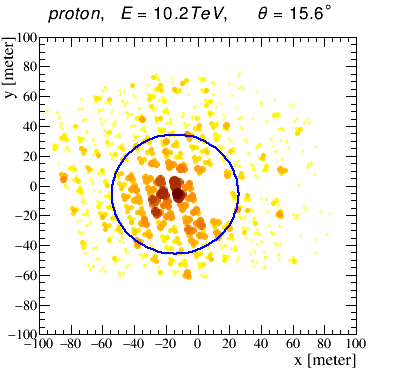}
  \caption{}
  \label{fig:pevent}
\end{subfigure} \\

\begin{subfigure}[t]{0.45\textwidth}
  \centering
  \includegraphics[width=1\linewidth]{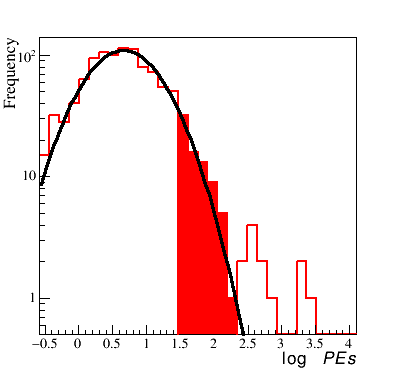}
  \caption{}
  \label{fig:gcharge}
\end{subfigure}
~
\begin{subfigure}[t]{0.45\textwidth}
  \centering
  \includegraphics[width=1\linewidth]{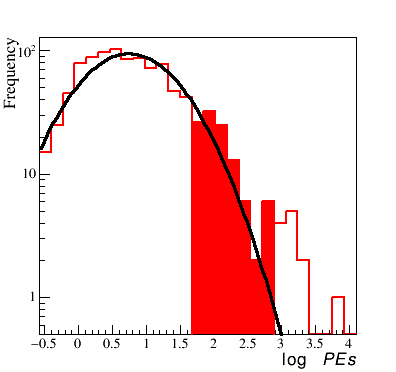}
  \caption{}
  \label{fig:pcharge}
\end{subfigure}

\caption{Comparison of \g (a) and proton (b) events of similar energy in HAWC. The saturation 
	and size of the dots is proportional to the signal strength at that location. Note the \g event's 
	well-defined core region and smooth lateral distribution beyond. Despite having a core region, 
	the proton event is relatively dispersed and scattered. The blue circle corresponds to the $40$~m
	radius used in the compactness method described in section 3.1. The signal distributions (in units of 
	photoelectrons, or PEs) for each shower are show in (c) and (d). The pair compactness hit 
	selection, $S$ (eq. 3.2), corresponds to the shaded region. The largest signals corresponding 
	to the core region are located beyond $S$.
	}
\label{fig:eventcomparison}
\end{figure}

Gamma-ray induced showers are nearly purely electromagnetic, which develop primarily
via pair-production and bremsstrahlung. Such showers are characterized by compact, 
smooth lateral shower profiles. Hadronic showers are dominated by large fluctuations caused by 
nuclear interactions, and subsequent generation and decay of pions. These sub-showers 
and further production of muons provide more transverse momentum as compared to \g-ray 
showers. This results in hadronic showers generally having larger signals farther from the shower's 
principal axis, or core (fig. \ref{fig:gevent}, \ref{fig:pevent}). We exploit this 
difference to suppress the hadronic background.

\section{The HAWC Observatory} \label{hawcobs}

The High-Altitude Water Cherenkov Observatory, or HAWC, is an air-shower array
optimized for the detection of gamma-ray showers. HAWC is located at $4100$~m above 
sea level on the slopes of Sierra Negra, Mexico, and comprises a $22,000~\rm{m^{2}}$ densely 
packed array of 300 water Cherenkov detectors (WCDs).  Each WCD is a light-tight 
steel tank $4.5$~m high  and $7.3$~m in diameter containing $200$~kL of purified water 
and four upward-facing photomultiplier tubes (PMTs). In its now completed configuration, 
HAWC is most sensitive to high-energy \g-rays and cosmic-rays in the energy 
range $100$~GeV -- $100$~TeV.

When an incident primary particle interacts in the Earth's atmosphere, 
a subsequent air-shower develops, containing many lower-energy
charged particles and \g-rays. As the secondary particles pass through the 
WCDs, they emit Cherenkov light, which is then detected by the PMTs. 
By measuring the timing, charge, and spatial information of the triggered PMTs in an air-shower, 
the arrival direction, energy, and type of the primary particle can be identified.

As shown here, topological event information can be used to discriminate the large background of
hadronic cosmic-ray showers ($10$~kHz event rate) from the electromagnetic air-showers 
produced by \g-rays. Suppressing this background, we can conduct a variety of studies including 
producing sky maps of extended and point-like \g-ray sources \cite{Ayala:2015, Hui:2015}, 
performing dark matter searches \cite{Harding:2015}, and measuring the \eeflux flux \cite{Benzvi:2015}. 
While this study illustrates one such \gh separation technique, another method in development 
can be found in \cite{Capistran:2015}. Further details on the operation of the HAWC detector and 
the event reconstruction are given in \cite{Smith:2015}. We use the 250 tank configuration of the detector 
(HAWC-250) for this study.

\section{Gamma/Hadron Separation Methods}

\subsection{Compactness}

One of the simplest \gh separation techniques used in HAWC analysis
is $compactness$ \cite{Smith:2015, Abeysekara:2013tza}. The method involves identifying the largest PMT signal, or hit, beyond $40$~m from the fit shower core ($CxPE40$), and dividing the total number of triggered 
PMTs ($nHit$) by this value:

\begin{equation} \label{eq:compact}
compactness=\frac{\text{nHit}}{\text{CxPE40}},
\end{equation}
where $CxPE40$ is measured in photoelectrons (PEs).

This measure has the intended effect of comparing detected air-shower events 
with similar sizes ($nHit$) and qualifying their gamma- or hadronic-like 
nature by exploiting fluctuations well outside the core region. As illustrated in 
fig.~\ref{fig:gevent}, \ref{fig:pevent}, the choice of $40$~m ensures that the 
core region, and its high concentration of large hits, is excluded from $CxPE40$. 
Hadronic showers, which tend to exhibit larger $CxPE40$ values than gamma showers, 
will be characterized by lower values of compactness than \g showers. 
Compactness can only be calculated when a fit to the core position has been successful.

\subsection{Pairwise Compactness}

The $pair$ $compactness$ method is an extension of compactness in that it involves 
measuring the spatial connectivity of a selection of hits. Where 
compactness identifies the relationship of the largest hit beyond $40$~m from the core, 
the pair compactness algorithm identifies the mean charge and pairwise spatial separation 
for a subset of hits.

The hit selection is made by observing that the distribution of the logarithm of the charges, \{$q$\}, in detected air-showers 
follows a nearly Gaussian distribution (fig.~\ref{fig:gcharge}, \ref{fig:pcharge}).
Using the Freedman-Diaconis Rule for optimal binning \cite{FDR:1981}, we fit a Gaussian to \{$\log q$\} 
and identify the mean and standard deviation ($\bar{q},\sigma_{q}$). If the fit fails, we simply extract $\bar{q}$ and 
$\sigma_{q}$ from the binned distribution. Those hits within the range $1.5\sigma$ and $3\sigma$ from $\bar{q}$ comprise 
the $n$ elements of the pair compactness hit selection, $S$:
% As the core region of events of any species 
%generally is characterized by large signals, and the small signals originate from the electromagnetic component of showers, 
%we use the shape of the charge distribution to define a subset of hits. 
%Those hits contained in the core region or the tail region 
%of the shower are excluded by selecting only hits within the range $1.5\sigma$ and $3\sigma$ from $\bar{q}$. 
%The hits in this charge range comprise the $n$ elements of the pair compactness hit selection, $S$:

\begin{equation} \label{eq:hit_selection}
	S=\{ i | \bar{q}+1.5\sigma_{q} < q_{i} < \bar{q}+3\sigma_{q}\},
\end{equation}
\begin{equation} \label{eq:hit_n}
	n=\{i | q_{i} \in S\}.
\end{equation}
The mean charge of this sub-range is calculated and identified as $q_{pc}$,
\begin{equation} \label{eq:qpc}
	q_{pc}=\frac{1}{n}\sum q_{i}.
\end{equation}
The spatial relationship is quantified as the weighted mean of the separation of hit pairs
in the hit selection, $r_{pc}$ (eq. \ref{eq:rpc}), where the weight assigned to each pair is the 
larger charge of the pair (eq. \ref{eq:wrpc}).
%In order to cap the number of calculations in computing hit pair distances, a maximum $n$ of 80 hits has been chosen (eq. \ref{eq:hit_n}).
\begin{equation} \label{eq:rpc}
		r_{pc}=\frac{1}{\sum\limits_{i \neq j}^{} w_{ij}} \sum \limits_{i \neq j} w_{ij} | r_{i} - r_{j} | ,
\end{equation}
where
\begin{equation} \label{eq:wrpc}
		w_{ij} = \max{q_{i}}{q_{j}}.
\end{equation}

\begin{figure}[H]
\centering
\begin{subfigure}[t]{0.49\textwidth}
  \centering
  \includegraphics[width=1\linewidth]{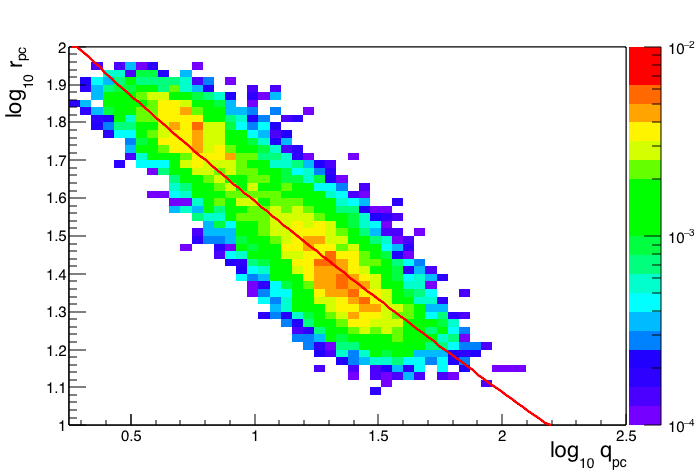}
  \caption{}
%  \label{fig:gevent}
\end{subfigure}
~
\begin{subfigure}[t]{0.49\textwidth}
  \centering
  \includegraphics[width=1\linewidth]{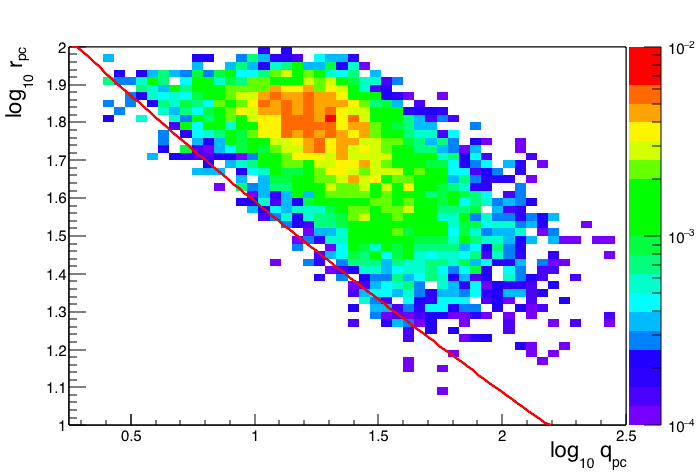}
  \caption{}
%  \label{fig:pevent}
\end{subfigure}
\caption{Normalized pair compactness distributions for gammas (a) and protons (b) for analysis bin 4, as defined in
table 1. The curve which maximizes the Q-factor (section 4.1) is shown in red.}
\label{fig:pc_dists}
\end{figure}

Similar to compactness, the pairwise compactness method aims to compare showers of similar size, 
but does so with the mean charge $q_{pc}$ of $S$ in lieu of $nHit$. Thus, the physical size of a shower 
represented by the mean separation distance, $r_{pc}$, is compared to showers of similar $q_{pc}$. 
In general, we expect hadronic showers to have greater $r_{pc}$ for a given $q_{pc}$, compared to \g showers. 
Based on the distribution of $q_{pc}$ and $r_{pc}$ for \g showers (fig. \ref{fig:pc_dists}), we make a fit of the form 
\begin{equation} \label{eq:fit_func}
  y = a+\frac{b}{x-c}+d \cdot x,
\end{equation}
then cut away all events that fall in the range $\log r_{pc} > y$, as shown in fig. \ref{fig:pc_dists}. 
In this manner we identify \g-like events with the pair compactness technique.

%Clearly, the pair compactness method is independent of whether the core was or was not fit successfully. 
%And since the number of hits in the hit selection, $S$, is constrained to the range $n<80$ (eq. \ref{eq:hit_n}), 
%pair compactness can be calculated at little computational cost in the event reconstruction for all 
%triggered showers, regardless of the core fit status.

\section{Comparison of Methods}

\subsection{Simulation of HAWC} \label{sec:sim}
To compare the performance of the separation methods, we simulate the
HAWC detector response to \g and hadronic air-showers. For an incident primary particle, 
the air-shower produced when it interacts in the atmosphere is simulated using the CORSIKA 
\cite{Heck:1998vt} package, with FLUKA \cite{Battistoni:2007, Ferrari:2005}, and QGSJET II 
\cite{Agostinelli:2002hh} packages as the particle physics interaction models. The HAWC detector 
response to the secondary particles is simulated using GEANT 4 configured to the HAWC 
detector configuration, tanks, and photomultipliers. For this study, simulation of gamma, proton, $^{4}He$, 
$^{12}C$, $^{16}O$, $^{20}Ne$, $^{24}Mg$, $^{28}Si$, and $^{56}Fe$ primaries was used to 
optimize and subsequently compare the separation capabilities of compactness and pair compactness.

%For this study, simulated pair compactness distributions were only available for proton and $^{4}He$ 
%events, while compactness has been optimized previously with additional simulation of $^{12}C$, 
%$^{16}O$, $^{20}Ne$, $^{24}Mg$, $^{28}Si$, and $^{56}Fe$ primaries. As protons and $^{4}He$ are 
%the dominant component of the cosmic-ray spectrum in the energy range defined in section \ref{hawcobs}, we 
%expect that this should be sufficient to estimate the separation power of pair compactness to 
%background cosmic rays.
%Details on the HAWC simulation framework can be found in refs. \cite{Smith:2015}.

We quantify the performance of the \gh rejection methods by calculating its $Q$-factor as defined here:
\begin{equation} \label{eq:qfactor}
  Q=\epsilon_\gamma/\sqrt{\epsilon_\text{CR}} ,
\end{equation}
where $\epsilon_\gamma$ and $\epsilon_\text{CR}$ are the efficiency to identify simulated 
gamma-rays and cosmic rays, respectively. In this study, we adjust the cut values
of the methods scanning for the maximal $Q$-factor. We also require $\epsilon_\gamma > 0.5$ 
to retain as many signal events as possible after cuts.

\subsection{Analysis Bins}
The current HAWC analysis defines a set of $nHit$ ranges in which cuts to compactness are optimized 
for the detection of the simulated transit of a Crab-like source. Included in the optimization of $Q$-factors for 
compactness are further cuts to goodness-of-fit results from the core and angular reconstruction. These additional
cuts are not included in the optimization of $Q$ for pair compactness which is done in the $nHit$ ranges defined in 
table~\ref{tab:bin_tab}. For each bin, we calculate the $Q$-factors for compactness and pair compactness, 
allowing a comparison of the two methods.

\begin{table}[H]
	\begin{center}
	\begin{tabular}{| l | c | r |}
	\hline
	Bin & $nHit$ Range \\	\hline
	0 & 38 - 46 \\	\hline
	1 & 47 - 70 \\	\hline
	2 & 71 - 111 \\	\hline
	3 & 112 - 171 \\	\hline
	4 & 172 - 253 \\	\hline
	5 & 254 - 353 \\	\hline
	6 & 354 - 466 \\	\hline
	7 & 467 - 580 \\	\hline
	8 & 581 - 679 \\	\hline
	9 & 680 - 865 \\	\hline
	\end{tabular}
	\end{center}
	\caption{Ranges of $nHit$ defining analysis bins.} \label{tab:bin_tab}
\end{table}

\subsection{Comparison of Methods}
\begin{figure}[H]
	\includegraphics[width=\linewidth]{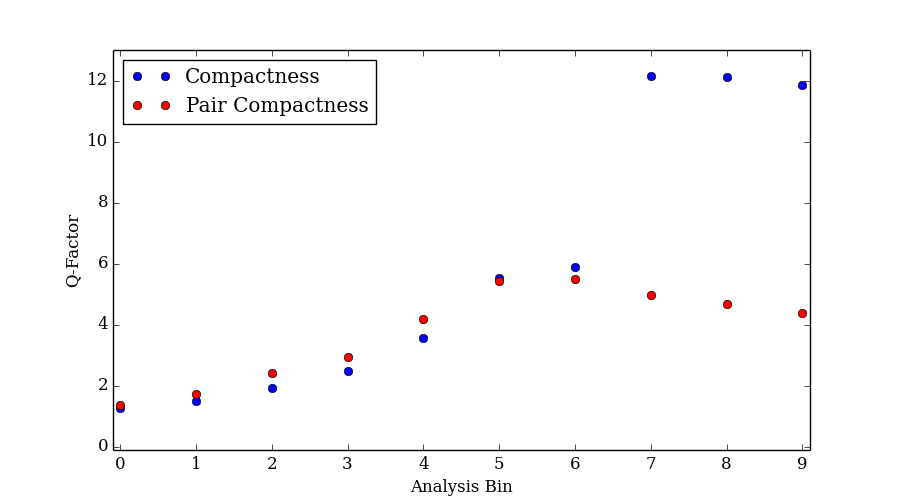}
	\caption{Q-factor comparison of compactness and pair compactness for HAWC-250 analysis bins. 
			In the higher analysis bins, the core region takes over the pair compactness calculation, 
			reducing its efficacy. 
			%Compactness continues to improve at the higher bins 
			%because they contain showers with larger $nHit$, thus the core fitter uncertainty 
			%is much lower, and thus compactness more reliable.}
			}
	\label{fig:qvalues}
\end{figure}

Figure \ref{fig:qvalues} shows the $Q$-factor comparison of compactness and pair compactness for each
analysis bin. Between bins 0 and 4, pair compactness has higher separation capability, with nearly matching $Q$-factors in bin 5. This may be due to the difficulty in event reconstruction, especially in the lowest $nHit$ bins. 
Both low-energy primaries landing on the array and high-energy primaries landing off the array produce 
lower-$nHit$ events which are difficult to fit to the steeply falling lateral distribution functions used in core fitting.

Beyond bin 6, compactness demonstrates significantly higher $Q$-factors. The decreased performance in pair 
compactness is caused by the saturation effect of the core region into $S$. Though the pair compactness algorithm 
attempts to exclude the core in the hit selection as shown in fig. \ref{fig:eventcomparison}, for high $nHit$ showers, 
the core region can be large enough in the number of hits to leak into $S$, dominating the weighting factors in the calculation of $r_{pc}$. The result is the convergence of the \g-ray and cosmic-ray pair compactness $r_{pc}$ 
distributions, i.e., the two types of showers begin to look similar at high $nHit$. Of course, the uncertainties 
in the event reconstruction drop as $nHit$ increases, contributing to the improved \gh separation from 
compactness in the higher analysis bins.

\section{Conclusion}
HAWC is a high-uptime air-shower array that is ideal for the study of both point-like and extended 
\g-ray sources. By employing algorithms that exploit topological differences between signal and background
events, we can increase the sensitivity of HAWC to these sources, as well as extend HAWC science capabilities 
to a variety of other studies. The pair compactness technique is a promising \gh separation method, especially for 
lower $nHit$ events, where there is an abundance of data and difficulties in fitting the shower parameters such as 
the core location required by compactness. Further improvements to the performance of pair compactness may be 
possible by identifying an improved hit selection, $S$, and by employing more sophisticated weighting schemes.

\section*{Acknowledgments}

\footnotesize{
We acknowledge the support from: the US National Science Foundation (NSF);
the US Department of Energy Office of High-Energy Physics;
the Laboratory Directed Research and Development (LDRD) program of
Los Alamos National Laboratory; Consejo Nacional de Ciencia y Tecnolog\'{\i}a (CONACyT),
Mexico (grants 260378, 55155, 105666, 122331, 132197, 167281, 167733);
Red de F\'{\i}sica de Altas Energ\'{\i}as, Mexico;
DGAPA-UNAM (grants IG100414-3, IN108713,  IN121309, IN115409, IN111315);
VIEP-BUAP (grant 161-EXC-2011);
the University of Wisconsin Alumni Research Foundation;
the Institute of Geophysics, Planetary Physics, and Signatures at Los Alamos National Laboratory;
the Luc Binette Foundation UNAM Postdoctoral Fellowship program.
}

\bibliography{icrc2015-0829}

\providecommand{\href}[2]{#2}\begingroup\raggedright\begin{thebibliography}{10}

\bibitem{Ayala:2015}
{\bf HAWC} Collaboration, H.~Ayala, {\it {Fermi Bubbles with HAWC}},  in {\em
  Proc. 34th ICRC}, (The Hague, The Netherlands), August, 2015.

\bibitem{Hui:2015}
{\bf HAWC} Collaboration, M.~Hui, {\it {HAWC Observation of Supernova Remnants
  and Pulsar Wind Nebulae}},  in {\em Proc. 34th ICRC}, (The Hague, The
  Netherlands), August, 2015.

\bibitem{Harding:2015}
{\bf HAWC} Collaboration, P.~Harding, {\it {Dark Matter Annihilation and Decay
  Searches with the High Altitude Water Cherenkov (HAWC) Observatory}},  in
  {\em Proc. 34th ICRC}, (The Hague, The Netherlands), August, 2015.

\bibitem{Benzvi:2015}
{\bf HAWC} Collaboration, S.~Benzvi, {\it {Measuring the e+ e- Flux Above 1 TeV
  with HAWC}},  in {\em Proc. 34th ICRC}, (The Hague, The Netherlands), August,
  2015.

\bibitem{Capistran:2015}
{\bf HAWC} Collaboration, T.~Capistran, {\it {New method for Gamma/Hadron
  separation in HAWC using neural networks}},  in {\em Proc. 34th ICRC}, (The
  Hague, The Netherlands), August, 2015.

\bibitem{Smith:2015}
{\bf HAWC} Collaboration, A.~Smith, {\it {HAWC: Design, Operation,
  Reconstruction and Analysis}},  in {\em Proc. 34th ICRC}, (The Hague, The
  Netherlands), August, 2015.

\bibitem{Abeysekara:2013tza}
A.~Abeysekara et~al., {\it {Sensitivity of the High Altitude Water Cherenkov
  Detector to Sources of Multi-TeV Gamma Rays}},  {\em Astropart.Phys.} {\bf
  50-52} (2013) 26--32, [\href{http://arxiv.org/abs/1306.5800}{{\tt
  arXiv:1306.5800}}].

\bibitem{FDR:1981}
P.~Freedman, D;~Diaconis, {\it {On the histogram as a density estimator: L2
  theory}},  {\em Probability Theory and Related Fields} {\bf 57 (4)} (1981)
  453--476.

\bibitem{Heck:1998vt}
D.~Heck, G.~Schatz, T.~Thouw, J.~Knapp, and J.~Capdevielle, {\it {CORSIKA: A
  Monte Carlo code to simulate extensive air showers}}, .

\bibitem{Battistoni:2007}
G.~Battistoni et~al., {\it {The FLUKA code: Description and benchmarking}},  in
  {\em {Proceedings of the Hadronic Shower Simulation Workshop 2006}}, vol.~896
  of {\em {AIP Conference Proceedings}}, p.~31, 2007.

\bibitem{Ferrari:2005}
A.~Ferrari, P.~Sala, A.~Fasso, and J.~Ranft, {\it {FLUKA: a multi-particle
  transport code}}, . {CERN-2005-10 (2005), INFN/TC\_05/11, SLAC-R-773}.

\bibitem{Agostinelli:2002hh}
{\bf GEANT4} Collaboration, S.~Agostinelli et~al., {\it {GEANT4: A Simulation
  Toolkit}},  {\em NIM} {\bf A506} (2003) 250--303.

\end{thebibliography}\endgroup

\end{document}